# The relationship between the radio core-dominance parameter and spectral index in different classes of extragalactic radio sources (*II*)

Zhi-Yuan Pei[1,2,3,4], Jun-Hui Fan[y,1,2], Denis Bastieri[1,3,4], Utane Sawangwit[5] and Jiang-He Yang[6]

[1] Center for Astrophysics, Guangzhou University, Guangzhou 510006, China; *mattpui@e.gzhu.edu.cn*

[2] Astronomy Science and Technology Research Laboratory of Department of Education of Guangdong Province, Guangzhou 510006, China

[3] Dipartimento di Fisica e Astronomia "G. Galilei", Universita` di Padova, I-35131 Padova, Italy

[4] Istituto Nazionale di Fisica Nucleare, Sezione di Padova, I-35131 Padova, Italy

[5] National Astronomical Research Institute of Thailand (NARIT), Chiang Mai, 50200, Thailand

[6] Department of Physics and Electronics Science, Hunan University of Arts and Science, Changde 415000, China

**Abstract** Active galactic nuclei (AGNs) can be divided into two major classes, namely radio loud and radio quiet AGNs. A small subset of the radio-loud AGNs is called blazars, which are believed to be unified with Fanaroff & Riley type I/II (FRI/II) radio galaxies. Following our previous work (Fan et al.2011), we present a sample of 2400 sources with measured radio flux densities of the core and extended components. The sample contains 250 BL Lacs, 520 quasars, 175 Seyferts, 1178 galaxies, 153 Fanaroff & Riley type I and type II galaxies and 104 unidentified sources. We then calculate the radio core-dominance parameters and spectral indices and study their relationship. Our analysis shows that the core-dominance parameters and spectral indices are quite different for different types of sources. We also confirm that the correlation between core-dominance parameter and spectral index exists for a large sample presented in this work.

## 1 INTRODUCTION

Active galactic nuclei (AGNs) are interesting extragalactic sources. Understanding these objects requires extensive knowledge in many different areas: accretion disks, the physics of dust and ionized gas, astronomical spectroscopy, star formation, and the cosmological evolution of galaxies and super massive black holes. Among AGNs, roughly 85% are radio-quiet AGNs, with radio loudness ratio, $B = \log \frac{F_{5\,\rm GHz}}{F_{2500\mathring{A}}} < 1.0$ and the remaining ~ 15% are radio-loud AGNs (Fan2005). A small subset of the radio-loud AGNs show frequent flux variability over the entire electromagnetic spectrum and strongly polarized emission. These objects are known as blazars.

[5] Corresponding author: fjh@gzhu.edu.cn

In addition, blazars are characterized by superluminal motions in their radio components, strong $\gamma$-ray emission etc. (see Abdo et al.2010;Aller et al.2003;Andruchow et al.2005;Cellone et al.2007;Fan et al.1997;Fan2005;Lin & Fan2018;Romero et al.2000,2002;Wills et al.1992;Xie et al.2005;Yang et al.2018a,b;Zhang & Fan2008). There are two subclasses of blazars, namely flat spectrum radio quasar (FSRQ) and BL Lacertae object (BL Lac), with the former showing strong emission line features and the latter showing very weak or no emission line at all. According to the peak frequency of the first bump in SEDs, $v_p$, BL Lacs are also divided into high-peak BL Ls (HBLs), intermediate-peak BL Lacs (IBLs), low-peak BL Lacs (LBLs). Nieppola et al.(2006) analysed spectral energy distribution, SEDs, for 308 blazars and set the boundaries at $\log v_p > 16.5$ for HBLs, $14.5 < \log v_p < 16.5$ for IBLs and $\log v_p < 14.5$ for LBLs. Abdo et al.(2010) analysed SEDs for 48 Fermi blazars and extended the definition to all types of non-thermal dominated AGNs using new acronyms: high-synchrotron-peaked blazars (HSP: $\log v_p > 15$), intermediate-synchrotron-peaked blazars (ISP: $14 < \log v_p < 15$) and low-synchrotron-peaked blazars (LSP: $\log v_p < 14$). In 2016, our group compiled multi-wavelength data for a sample of 1425 Fermi blazars, significantly larger sample than that studied by Abdo et al.(2010), using a parabolic function to fit their SEDs. Adopting "Bayesian classification", we found refined boundaries for HSP, ISP and LSP blazars at $\log v_p > 15.3$, $14.0 < \log v_p \leq 15.3$ and $\log v_p \leq 14.0$, respectively (Fan et al.2016).

Blazars are strong $\gamma$-ray emitters. Many studies highlight the role of beaming in producing strong $\gamma$-ray emission. In addition, their emission mechanism, central structure, and evolution among different types of subclasses of AGNs are still under debate (e.g., Abdo et al.2009;Bhattacharya et al.2016;Chen et al. 2013;Fan et al.2010,2013;Lin & Fan2018;Liodakis2018;Mao et al.2005;Romero et al.2002;Xue et al.2017;Zhang et al.2015). The standard model of AGNs predicts that FRIs are the parent population of BL Lacs while the parent population of FSRQs are FRIIs (Urry & Padovani1995, also see Fan et al.2011).

After the launch of Fermi gamma-ray telescope, many additional sources were $\gamma$-ray detected, providing a good opportunity to study the high energy astrophysics and gamma-ray emission mechanisms. For example, the latest Fermi-LAT 4-year point source catalogue, 3FGL, contains 1444 blazars (Ackermann et al.2015). Recently, we compiled a sample of 1335 blazars with available core-dominance parameter $R$, consisting of 169 $\gamma$-ray sources (Fermi blazars), and compared the $\log R$ values of the Fermi and non-Fermi blazars (Pei et al.2016). We found the average of $\log R$ of Fermi blazars to be far greater than that of non-Fermi ones. Therefore, the $\gamma$-ray blazars are more core-dominated and beamed than non-Fermi ones.

In a relativistic beaming model, the emission is assumed to be produced by two components, namely the beamed and the unbeamed ones, or core and extended ones. Then, the observed total emission, $S^{ob}$, is the sum of the beamed, $S^{ob}_{core}$, and unbeamed, $S_{ext.}$, emission, use $S^{ob} = S_{ext.} + S^{ob}_{core} = (1 + f^{\delta p})S_{ext.}$, where $f = \frac{S^{in}_{core}}{S_{ext.}}$, $S^{in}_{core}$ is the de-beamed emission in the co-moving frame, $\delta$ is a Doppler factor, $p = \alpha + 2$ (for a continuos jet case) or $p = \alpha + 3$ (for a moving sphere case), and $\alpha$ is the spectral index ($S_\nu \propto \nu^{-\alpha}$). The ratio, $R$, of the two components is the core-dominance parameter. Some authors use the ratio of flux densities while others use the ratio of luminosities to quantify the parameter. Namely, $R = S_{core}/S_{ext.}$ or $R = L_{\text{core}}/L_{\text{ext.}}$, where $S_{\text{core}}$ or $L_{\text{core}}$ stands for core emission while $S_{\text{ext.}}$ or $L_{\text{ext.}}$ for extended emission (seeFan & Zhang2003;Fan et al.2011and references therein).

In 2011, we compiled a catalogue of 1223 AGNs, calculated their core-dominance parameters and investigated the correlation between the radio spectral index and core-dominance parameter (Fan et al. 2011).

In this work, followingFan et al.(2011), we collect a new sample of radio sources, which were not included inFan et al.(2011), calculate the core-dominance parameters and the radio spectral indices, and then revisit the correlation among them. In this sense, we extend the AGN sample with regard to the available core-dominance parameter log $R$ and make additional discussions and re-examine the conclusions drawn in our previous work (Fan et al.2011). Our data are taken from the NASA/IPAC EXTRAGALACTIC DATABASE (http://ned.ipac.caltech.edu/forms/byname.html), SIMBAD Astronomical Database (http://simbad.u-strasbg.fr/simbad/) and Roma BZCAT (http://www.asdc.asi.it/bzcat/), from these, we calculate the core-dominance parameters and spectral indices of 2400 sources with available radio data. In section 2, we will present the results; some discussions are given in section 3. We then conclude and summarise our findings in the final section.

Throughout this paper, without loss of generality, we take $\Lambda CDM$ model, with $\Omega_\Lambda$ c 0.73, $\Omega_M$ c 0.27, and $H_0$ c 73 km s$^{-1}$ Mpc$^{-1}$.

## 2 SAMPLE AND RESULTS

### 2.1 Sample and Calculations

In order to calculate the radio core-dominance parameter and discuss its properties, we compiled a list of relevant data from the literature. In general, the observations were performed at different frequencies by various authors and studies. However, most of these data are at 5 GHz, we therefore transformed the data, given in the literature at other frequencies ($\nu$), to 5 GHz using the assumption that (Fan et al.2011;Pei et al. 2016)

$$S^{5\,\text{GHz}}_{\text{core}} = S^{\nu,\text{obs}}_{\text{core}}, \quad S^{5\,\text{GHz}}_{\text{ext.}} = S^{\nu,\text{obs}}_{\text{ext.}} \left(\frac{\nu}{5\,\text{GHz}}\right)^{\alpha_{\text{ext.}}}, \quad (1)$$

then the flux densities are K-corrected, and the core-dominance parameters are finally calculated, using the expression

$$R = \left(\frac{S_{\text{core}}}{S_{\text{ext.}}}\right)(1+z)^{\alpha_{\text{core}} - \alpha_{\text{ext.}}}, \quad (2)$$

In our calculation, we adopted $\alpha_{\text{ext.}}$ (or $\alpha_{\text{unb}}$) = 0.75 and $\alpha_{\text{core}}$ (or $\alpha_j$) = 0 (Fan et al.2011). All data in the table are flux densities, some of them where luminosities that we transform, if necessary, at 5 GHz. Then we calculated the core-dominance parameter as $\log R = \log \frac{L_{\text{core}}}{L_{\text{ext.}}}$. For flux density data, we also calculate the luminosity using $L_\nu = 4\pi d_L^2 S_\nu$, where $d_L$ stands for a luminosity distance, defined by $d_L = (1+z)\frac{c}{H_0}\int_1^{1+z}\frac{1}{\sqrt{\Omega_M x^3 + 1 - \Omega_M}}dx$. The data are mainly at 1.4 and 5 GHz obtained from the literature, we then calculated the spectral indices, $\alpha$ where $S_\nu \propto \nu^{-\alpha}$. If a source has no measured redshift, then the average value of the corresponding group was adopted, only to be used to calculate the core-dominance parameters and luminosities. We evaluate the characteristics of 2400 sources and checked their identification using the NED and Roma BZCAT. The former gives the basic and initial identification while the latter gives the classifications either as a BL Lac or a quasar. Furthermore, if a source is identified as "QSO" in NED, we then check whether it is identified as BL Lac, quasar or uncertain type (herein using "unidentified") in BZCAT. If a source is identified as "G", we then only use FRI and FRII to identify the source. For the classification of Seyfert galaxies, we use "Seyfert". For those sources that are not identified as FRI, FRII or Seyfert, we use "galaxy" to label them. If the object has no identification in NED, it is also labeled as "unidentified".

In total, our literature survey found 250 BL Lacs, 520 quasars, 175 Seyferts, 1178 galaxies, 153 FRIs & FRIIs and 104 unidentified sources. To obtain the core flux $S_{core}$, we search through a large number of references and databases containing core emission, crosscheck these sources with the catalogue given by Fan et al.(2011), and choose the sources that were not included in Fan et al.(2011). For the case that the core flux given by the literature is not at 5 GHz, we transform it to 5 GHz with respect to Equation (1), assuming that the core emission at different frequency are the same for one source since $\alpha_{core} \sim 0$ is adopted for the core spectral index. For the extended and total emission, we have two methods: (i) if the flux density is provided by the literature, we use this value (if the flux given is not at 5 GHz, we convert it to 5 GHz by Equation (1)), and then we obtained the total emission at 5 GHz ($S_{Total} = S_{core} + S_{ext.}$), otherwise (ii) we query the SED for the total flux of this source at 5 GHz, and obtain the extended flux ($S_{ext.} = S_{Total} - S_{core}$). We then calculate the K-corrected core dominance parameter, $R$.

Our sample was obtained from a variety of observational campaigns from different facilities, such as VLA, VLBI, MOJAVE, TANAMI, etc, which are affected by different systematics. One may also consider that, since variability is a typical property of AGNs, different values may refer to different epochs. For this reason, as there could be different flux densities reported by different instruments, we adopt the maximum value at the highest resolution.

The data and their corresponding references are shown in Table 1. The complete Table for the whole sample is attached as an online material of this paper. Col. 1 gives the source names; Col. 2 classification (B: BL Lac, Q: quasar; S: Seyfert; G: galaxy; FRI: Fanaroff and Riley type I; FRII: Fanaroff and Riley type II; U: unidentified); Col. 3 redshift, $z$; Col. 4 frequency in GHz for emission; Col. 5 core-emission in mJy; Col. 6 extended emission in mJy and Col. 7 total emission in mJy; Col. 8 references for Col 5, Col. 6 and Col. 7; Col. 9 core-dominance parameter at 5 GHz, $\log R$; Col. 10 frequency in Ghz; Col. 11 total emission in mJy, data in Col. 10 and Col. 11 are from NED; Col. 12 the radio spectral index, $\alpha$ ($S_\nu \propto \nu^{-\alpha}$). Data in this table are taken from B06:Balmaverde et al.(2006); B11:Broderick & Fender(2011); C05: Capetti et al.(2005); C11:Caramete et al.(2011); C99:Cassaro et al.(1999); D12:Doi et al.(2012); D15:Duțan & Caramete(2015); D99:Dallacasa et al.(1999); DM14:Di Mauro et al.(2014); FXG06: Fan et al.(2006); G01:Giovannini et al.(2001); GG04:Giroletti et al.(2004); H12:Hlavacek-Larrondo et al.(2012); K04:Kellermann et al.(2004); K05:Kovalev et al.(2005); K10:Kharb et al.(2010); K96: Kollgaard et al.(1996); L01:Lister et al.(2001); L09:Lin et al.(2009); L09a:Lister et al.(2009); L11: Lal et al.(2011); LM93:Laurent-Muehleisen et al.(1993); LM97:Laurent-Muehleisen et al.(1997); M16: Marin & Antonucci(2016); M93:Morganti et al.(1993); M93a:Murphy et al.(1993); P03:Pollack et al. (2003); P14:Piner & Edwards(2014); P16:Panessa et al.(2016); P96:Perlman et al.(1996); S16:Smith

Table 1 Sample for the whole sources

| Name | Class | z | $v_1$ | $S_{core}$ | $S_{ext.}$ | $S_1^{Total}$ | Ref. | log R | $v_2$ | $S_2^{Total}$ | α |
|---|---|---|---|---|---|---|---|---|---|---|---|
| (1) | (2) | (3) | (4) | (5) | (6) | (7) | (8) | (9) | (10) | (11) | (12) |
| 0033+595 | B | 0.086 | 5.0 | 62 | 5 | 67 | PS96 | 1.03 | 1.4 | 148.3 | 0.64 |
| 0007+1708 | Q | 1.601 | 5.0 | 960 | 29 | 989 | LM97 | 0.74 | 2.7 | 910 | -0.14 |
| 0003+158 | S | 0.451 | 5.0 | 24 | 316 | 340 | Z11 | -0.85 | 2.7 | 885 | 1.00 |
| 0003+123 | G | 0.98 | 5.0 | 156 | 53 | 209 | LM97 | 0.28 | 1.4 | 224 | 0.05 |
| 0315+41 | FRI | 0.026 | 5.0 | 40 | 3490 | 3530 | M13 | -1.90 | 2.7 | 4970 | 0.56 |
| 1136-163 | U | 0.550 | 5.0 | 34 | 218 | 252 | LM97 | -0.58 | 1.4 | 160.52 | 1.11 |
| … | … | … | … | … | … | … | … | … | … | … | … |

Notes: In the Table, Col. 1 gives the name of the source; Col. 2 classification (B: BL Lac, Q: quasar; S: Seyfert; G: galaxy; FRI: Fanaroff and Riley type I; FRII: Fanaroff and Riley type II; U: undentified); Col. 3 redshift, z; Col. 4 frequency in GHz for emission; Col. 5 core-emission in mJy; Col. 6 extended emission in mJy and Col. 7 total emission in mJy; Col. 8 references for Col 5, Col. 6 and Col. 7; Col. 9 core-dominance parameter at 5 Ghz, log R; Col. 10 frequency in GHz; Col. 11 total emission in mJy, data in Col. 10 and Col. 11 are from NED; Col. 12 the radio spectral index, α ($S_\nu \propto \nu^{-\alpha}$).

et al.(2016); T96:Taylor et al.(1996); W06:Wang et al.(2006a); W07:Wu et al.(2007); W13:Wilkes et al.(2013); W14:Wu et al.(2014); Y12:Yuan & Wang(2012).

Data are K-corrected using $S = S^{ob}(1+z)^{(\alpha-1)}$ ($S \propto \nu^{-\alpha}$) with $\alpha_{core} = 0$ and $\alpha_{ext.} = 0.75$ adopted. Then, we calculate the core-dominance parameter, R, using $\log R = \log \frac{S_{core}}{S_{ext.}}$ (Col. (9)). For the radio spectral index, we calculate it from $\alpha = -\frac{\log(S_1^{Total}/S_2^{Total})}{\log(\nu_2/\nu_1)}$ and list them in Col. (12).

## 2.2 Estimated parameters

For the whole sample, we can calculate the average value for the core-dominance parameter (log R).

We found that log R are in the range from −3.50 to 3.96 with an average value of $(\log R)|_{Total}$ = −0.34 ± 1.06 for the whole 2400 sources; from −1.34 to 3.35 with an average value of $(\log R)|_{BL\ Lac}$ = 0.55 ± 0.91 for 250 BL Lacs; from −3.08 to 3.83 with an average value of $(\log R)|_{quasar}$ = 0.24 ± 1.03 for 521 quasars; from −2.89 to 0.56 with an average value of $(\log R)|_{Seyfert}$ = −0.37 ± 0.61 for 175 Seyfert galaxies; from −3.50 to 3.96 with an average value of $(\log R)|_{galaxy}$ = −0.67 ± 0.96 for 1178 galaxies; from −3.25 to 0.72 with an average value of $(\log R)|_{FRI}$ = −0.97 ± 0.84 for 46 FR type I radio galaxies; from −3.25 to 0.72 with an average value of $(\log R)|_{FRII}$ = −1.38 ± 0.54 for 107 FR type II radio galaxies; from −1.36 to 1.67 with an average value of $(\log R)|_{unidentifted}$ = −0.26 ± 0.72 for 104 unidentified sources (see Table 2).

Therefore, we found that the average core-dominance parameters for the sources follow the relation: $(\log R)|_{BL\ Lac}$ > $(\log R)|_{quasar}$ > $(\log R)|_{Seyfert}$ > $(\log R)|_{galaxy}$ > $(\log R)|_{FRI}$ > $(\log R)|_{FRII}$. Using the core-dominance parameter, blazars appear to be the most core-dominated population of AGNs.

For the radio spectral index (α), from −2.42 to 2.42 with an average value of $(\alpha)|_{Total}$ = 0.41±0.65 for the whole sample; from −1.88 to 1.48 with an average value of $(\alpha)|_{BL\ Lac}$ = 0.22±0.50 for BL Lacs; from −1.66 to 2.08 with an average value of $(\alpha)|_{quasar}$ = 0.15 ± 0.61 for quasars; from −2.42 to 1.84 with an average value of $(\alpha)|_{Seyfert}$ = 0.43±0.95 for Seyfert galaxies; from −2.05 to 1.88 with an average value of $(\alpha)|_{galaxy}$ = 0.57±0.54 for galaxies; from −1.70 to 1.84 with an average value of $(\alpha)|_{FRI}$ = 0.69±0.49 for FR type I radio galaxies; from $-1.43$ to $1.57$ with an average value of $(\alpha)|_{\rm FRII} = 0.60 \pm 0.58$ for FR type II radio galaxies; from $-2.36$ to $2.42$ with an average value of $(\alpha)|_{\rm unidentified} = 0.00 \pm 1.35$ for unidentified sources (see Table 2).

**Table 2** Average values for the whole sample

| Sample | $N$ | $(\log R)$ | $(\alpha)$ | $(\log L_{\rm core})$ $({\rm W \cdot Hz^{-1}})$ | $(\log L_{\rm ext.})$ $({\rm W \cdot Hz^{-1}})$ |
|---|---|---|---|---|---|
| Total | 2400 | $-0.34 \pm 1.06$ | $0.41 \pm 0.66$ | $24.82 \pm 1.80$ | $25.23 \pm 1.72$ |
| BL Lac | 250 | $0.55 \pm 0.91$ | $0.22 \pm 0.50$ | $25.34 \pm 1.18$ | $24.93 \pm 1.32$ |
| Quasar | 520 | $0.24 \pm 1.03$ | $0.15 \pm 0.61$ | $26.44 \pm 1.28$ | $26.13 \pm 1.28$ |
| Seyfert | 175 | $-0.37 \pm 0.61$ | $0.43 \pm 0.95$ | $23.43 \pm 2.29$ | $23.77 \pm 2.30$ |
| Galaxy | 1178 | $-0.67 \pm 0.96$ | $0.57 \pm 0.54$ | $24.06 \pm 1.53$ | $24.97 \pm 1.83$ |
| FRI | 46 | $-0.97 \pm 0.84$ | $0.69 \pm 0.49$ | $22.99 \pm 0.77$ | $24.95 \pm 1.05$ |
| FRII | 107 | $-1.38 \pm 0.54$ | $0.60 \pm 0.58$ | $24.31 \pm 0.75$ | $26.29 \pm 1.23$ |
| Unidentified | 104 | $-0.26 \pm 0.72$ | $0.00 \pm 1.35$ | $\cdots$ | $\cdots$ |

We can find that the average radio spectral indices for the sources follows a trend: $(\alpha)|_{\rm FRI} > (\alpha)|_{\rm FRII} > (\alpha)|_{\rm galaxy} > (\alpha)|_{\rm Seyfert} > (\alpha)|_{\rm BL\ Lac} > (\alpha)|_{\rm quasar}$, which is basically opposite to the case in $\log R$. Thus, the mean value of radio spectral indices of quasars takes the minimum in all kinds of AGNs, indicating that the radio quasars population in our sample is dominated by FSRQs.

For the core luminosity $\log L_{\rm core}$ (W · Hz$^{-1}$), from $18.04$ to $29.98$ with an average value of $(\log L_{\rm core})|_{\rm Total} = 24.82 \pm 1.80$ for the whole sample; from $21.83$ to $28.07$ with an average value of $(\log L_{\rm core})|_{\rm BL\ Lac} = 25.34 \pm 1.18$ for BL Lacs; from $20.51$ to $28.69$ with an average value of $(\log L_{\rm core})|_{\rm quasar} = 26.44 \pm 1.28$ for quasars; from $19.58$ to $28.87$ with an average value of $(\log L_{\rm core})|_{\rm Seyfert} = 23.43 \pm 2.29$ for Seyfert galaxies; from $18.04$ to $29.98$ with an average value of $(\log L_{\rm core})|_{\rm galaxy} = 24.06 \pm 1.53$ for galaxies; from $20.90$ to $24.50$ with an average value of $(\log L_{\rm core})|_{\rm FRI} = 22.99 \pm 0.77$ for FR type I radio galaxies; from $23.08$ to $25.95$ with an average value of $(\log L_{\rm core})|_{\rm FRII} = 24.31 \pm 0.75$ for FR type II radio galaxies (also see Table 2).

For the extended luminosity $\log L_{\rm ext.}$ (W · Hz$^{-1}$), from $17.80$ to $30.88$ with an average value of $(\log L_{\rm ext.})|_{\rm Total} = 25.23 \pm 1.72$ for the whole sample; from $20.41$ to $29.47$ with an average value of $(\log L_{\rm ext.})|_{\rm BL\ Lac} = 24.93 \pm 1.32$ for BL Lacs; from $20.97$ to $29.76$ with an average value of $(\log L_{\rm ext.})|_{\rm quasar} = 26.13 \pm 1.28$ for quasars; from $20.41$ to $28.34$ with an average value of $(\log L_{\rm ext.})|_{\rm Seyfert} = 23.77 \pm 2.30$ for Seyfert galaxies; from $17.80$ to $30.88$ with an average value of $(\log L_{\rm ext.})|_{\rm galaxy} = 24.97 \pm 1.83$ for galaxies; from $22.85$ to $26.87$ with an average value of $(\log L_{\rm ext.})|_{\rm FRI} = 24.95 \pm 1.05$ for FR type I radio galaxies; from $23.05$ to $28.35$ with an average value of $(\log L_{\rm ext.})|_{\rm FRII} = 26.29 \pm 1.23$ for FR type II radio galaxies (also see Table 2).

### 2.3 Distributions of Core-Dominance Parameters and Spectral Indices

Figure 1 shows the distribution of core-dominance parameter, $\log R$ (a) and the cumulative probability (b) for the subclasses of our sample.

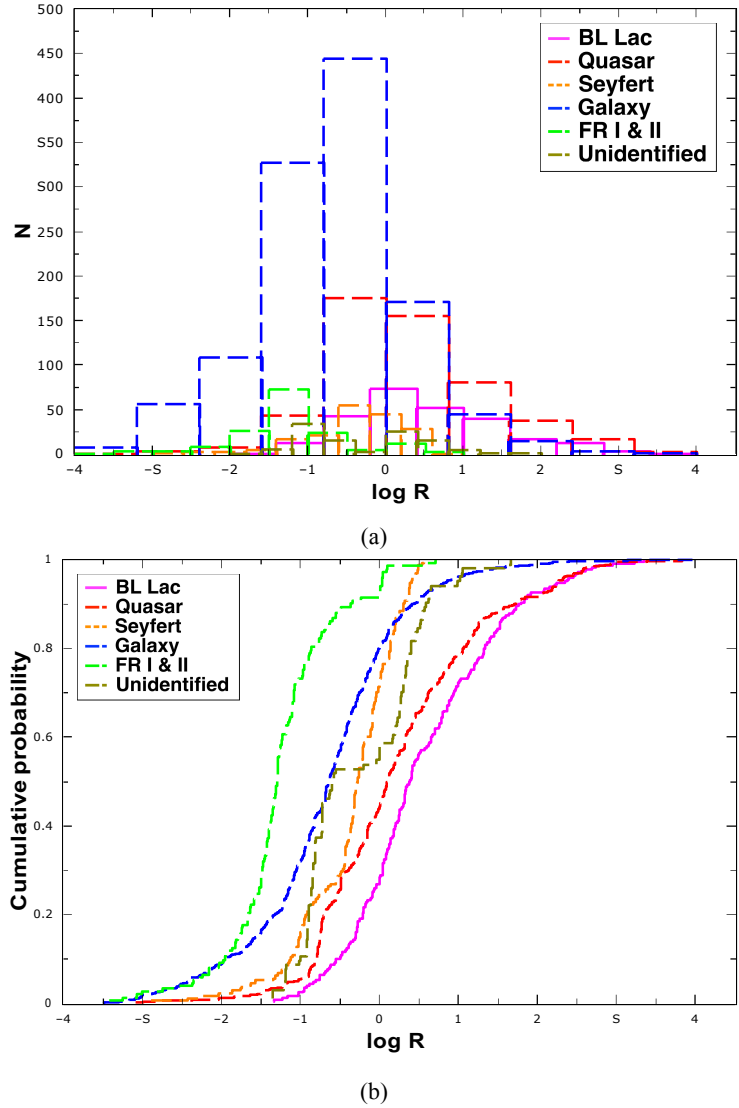

**Fig. 1** Distribution of core-dominance parameter, $\log R$ (a) and the cumulative probability (b) for the whole sample. In this plot, magenta solid line stands for BL Lacs, red dash line for quasars, orange dotted line for Seyferts, blue dash-dotted line for galaxies, green dash-dotted-dotted line for FR type I and II radio galaxies and dark yellow short dash line for unidentified sources.

A Kolmogorov-Smirnov (K-S) test rejects the hypothesis that BL Lacs and quasars have the same parent distribution at $p = 2.52 \times 10^{-5}$ ($d_{max} = 0.18$). Likewise, for BL Lacs and Seyferts, we have $p = 3.57 \times 10^{-18}$ ($d_{max} = 0.47$); for BL Lacs and galaxies, we have $p = 8.02 \times 10^{-20}$ ($d_{max} = 0.53$); for BL Lac and FRI galaxies, we have $p = 1.13 \times 10^{-13}$ ($d_{max} = 0.61$); for BL Lac and FRII galaxies, we have $p = 6.49 \times 10^{-23}$ ($d_{max} = 0.87$); for quasars and Seyferts, we have $p = 1.77 \times 10^{-13}$ ($d_{max} = 0.34$); for quasars and galaxies, we have $p = 5.14 \times 10^{-19}$ ($d_{max} = 0.36$); for quasar and FRI galaxies, we have $p = 3.24 \times 10^{-11}$ ($d_{max} = 0.53$); for quasar and FRII galaxies, we have $p = 1.76 \times 10^{-21}$ ($d_{max} = 0.81$) (see Table 3).

Figure 2 shows the distributions of radio spectral index, $\alpha$ (a) and the cumulative probability (b) for the subclasses.

**Table 3** Statistical results for core-dominance parameter (log $R$) in the whole sample

| Sample: A-B | $N_A$ | $N_B$ | $d_{max}$ | $p$ |
|---|---|---|---|---|
| BL Lac-Quasar | 250 | 520 | 0.18 | $2.52 \times 10^{-5}$ |
| BL Lac-Seyfert | 250 | 175 | 0.47 | $3.57 \times 10^{-18}$ |
| BL Lac-Galaxy | 250 | 1178 | 0.53 | $8.02 \times 10^{-20}$ |
| BL Lac-FRI | 250 | 46 | 0.61 | $1.13 \times 10^{-13}$ |
| BL Lac-FRII | 250 | 107 | 0.87 | $6.49 \times 10^{-23}$ |
| Quasar-Seyfert | 520 | 175 | 0.34 | $1.77 \times 10^{-13}$ |
| Quasar-Galaxy | 520 | 1178 | 0.36 | $5.14 \times 10^{-19}$ |
| Quasar-FRI | 520 | 46 | 0.53 | $3.24 \times 10^{-11}$ |
| Quasar-FRII | 520 | 107 | 0.81 | $1.76 \times 10^{-21}$ |

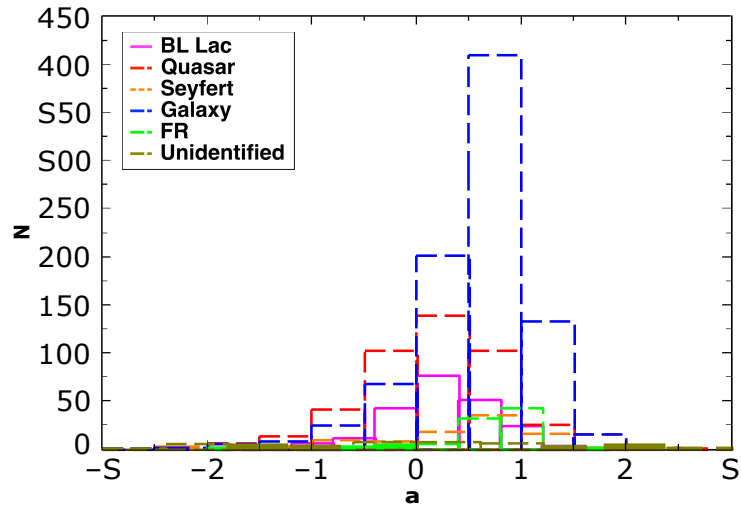

(a)

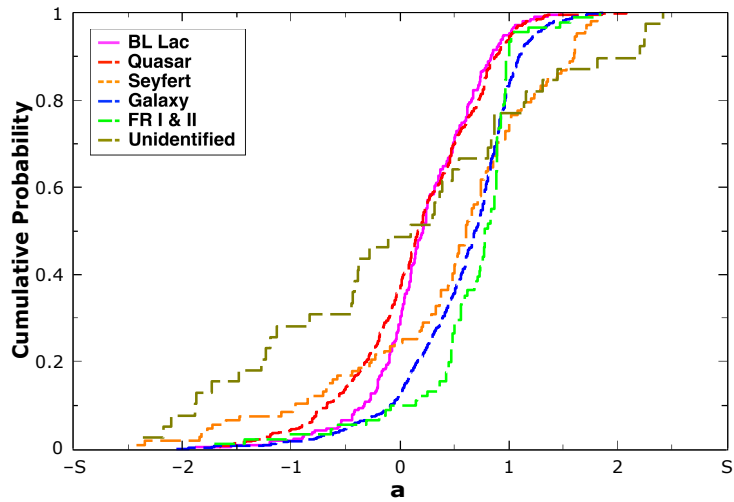

(b)

**Fig. 2** Distribution of radio spectral index, $\alpha$ (a) and the cumulative probability (b) for the whole sample. In this plot, magenta solid line stands for BL Lacs, red dash line for quasars, orange dotted line for Seyferts, blue dash-dotted line for galaxies, green dash-dotted-dotted line for FR type I and II radio galaxies and dark yellow short dash line for unidentified sources.

**Table 4** Statistical results for radio spectral index ($\alpha$) in the whole sample

| Sample: A-B | $N_A$ | $N_B$ | $d_{max}$ | $p$ |
|---|---|---|---|---|
| BL Lac-Quasar | 214 | 427 | 0.11 | 7.6 % |
| BL Lac-Seyfert | 214 | 107 | 0.30 | $1.26 \times 10^{-10}$ |
| BL Lac-Galaxy | 214 | 862 | 0.36 | $6.47 \times 10^{-13}$ |
| BL Lac-FRI | 214 | 42 | 0.55 | $3.15 \times 10^{-10}$ |
| BL Lac-FRII | 214 | 49 | 0.50 | $2.35 \times 10^{-9}$ |
| Quasar-Seyfert | 427 | 107 | 0.31 | $1.56 \times 10^{-7}$ |
| Quasar-Galaxy | 427 | 862 | 0.34 | $8.24 \times 10^{-11}$ |
| Quasar-FRI | 427 | 42 | 0.54 | $1.17 \times 10^{-10}$ |
| Quasar-FRII | 427 | 49 | 0.47 | $4.26 \times 10^{-9}$ |

**Table 5** Statistical results for extended-Luminosity ($\log L_{ext.}$) in the sample of BL Lacs, quasars, FRIs and FRIIs

| Sample: A-B | $N_A$ | $N_B$ | $d_{max}$ | $p$ |
|---|---|---|---|---|
| BL Lac-Quasar | 202 | 377 | 0.39 | $4.86 \times 10^{-12}$ |
| BL Lac-FRI | 202 | 46 | 0.16 | 27.5% |
| Quasar-FRII | 377 | 59 | 0.17 | 9.03% |

For radio spectral index $\alpha_R$, we obtain the following results: $p = 7.6\%(d_{max} = 0.11)$ for BL Lacs and quasars; $p = 1.26 \times 10^{-10}$ ($d_{max} = 0.30$) for BL Lacs and Seyferts; $p = 6.47 \times 10^{-13}$ ($d_{max} = 0.36$) for BL Lacs and galaxies; $p = 3.15 \times 10^{-10}$ ($d_{max} = 0.55$) for BL Lacs and FRIs; $p = 2.35 \times 10^{-9}$ ($d_{max} = 0.50$) for BL Lacs and FRIIs. Besides, $p = 1.56 \times 10^{-7}$($d_{max} = 0.31$) for quasars and Seyferts; $p = 8.24 \times 10^{-11}$ ($d_{max} = 0.34$) for quasars and galaxies; $p = 1.17 \times 10^{-10}$ ($d_{max} = 0.54$) for quasars and FRIs; $p = 4.26 \times 10^{-9}$ ($d_{max} = 0.47$) for quasars and FRIIs (see Table 4).

Through the K-S tests, we found that the distributions of $\log R$ and $\alpha$ in the various subclasses are almost significantly different, indicating that there are many different intrinsic properties among all the subclasses. However, considering the BL Lacs versus quasars with regard to $\alpha$, the result from K-S test shows that there is no significant difference (with chance probability of 7.6%), which implies that, as the two subclasses of blazars, they hold some similar inherent properties.

The distributions of extended luminosity, $\log L_{ext.}$ for BL Lacs, FRIs, FRIIs and quasars are shown in Figure 3. Our K-S test for the probability that any two samples may be drawn from one single parent sample found that: $p = 4.86 \times 10^{-12}$ ($d_{max} = 0.39$) for BL Lacs and quasars; $p = 27.5\%$ ($d_{max} = 0.16$) for BL Lacs and FRIs; $p = 9.03\%$ ($d_{max} = 0.17$) for quasars and FRIIs. (see Table 5).

In view of these K-S test results, we believe BL Lacs should be unified with FR type I galaxies and quasars should be unified with FR type II galaxies.

**2.4 Correlation Analysis**

We now turn our attention to linear correlation, if any, of the extended and core luminosity. In the two-component beaming model, the core emission are supposed to be the beamed component and the extended emission to be the unbeamed one. Because of the core-dominance parameter can take the indication of orientation, thus $\log R$ is also the indication of beaming effect. We use the extended luminosity at 5 GHz, $\log L_{\text{ext.}}$, to study the relationship between the beaming effect and unbeamed emission. Here, we K-corrected the flux densities, and then calculate the luminosity by $L_v = 4\pi d_L^2 S_v$, where $d_L$ is the luminosity distance. We found the relation that $\log L_{\text{ext.}} = (0.55 \pm 0.02) \log L_{\text{core}} + (11.70 \pm 0.53)$ with a correlation coefficient $r = 0.67$ and a chance probability of $p \sim 0$ for the whole sample as shown in Figure 4, which shows that the extended luminosity $\log L_{\text{ext.}}$ increases with increasing core luminosity $\log L_{\text{core}}$.

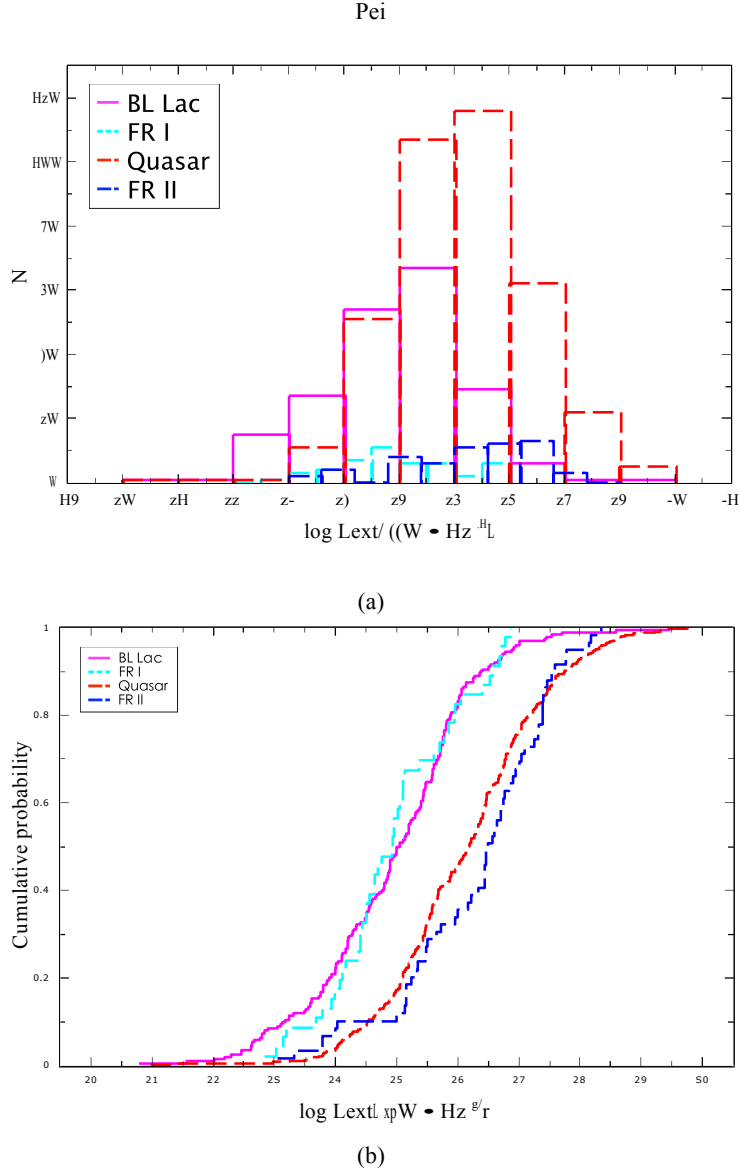

**Fig. 3** Distribution of extended luminosity, $\log L_{\text{ext.}}$ (a) and the cumulative probability (b) for the sample of BL Lacs, quasars, FR type I and type II. In this plot, magenta solid line stands for BL Lacs, red dash line for quasars, cyan dotted line for FRIs and blue dash-dotted line for FRIIs.

A correlation between core-dominance parameter and extended luminosity is found for the whole sample (see Figure 5(a)), which shows $\log L_{\text{ext.}} = -(0.30 \pm 0.04) \log R + (25.15 \pm 0.05)$ with a correlation coefficient $r = -0.21$ and a chance probability of $p = 1.89 \times 10^{-15}$. In this plot, the representations of all symbols are the same as in Figure 1. Figure 5(b) to Figure 5(f) show the correlations between $\log R$ and log $L_{ext.}$ for BL Lacs, quasars, Seyferts, galaxies and FRIs & FRIIs, respectively. For BL Lacs, we have log $L_{ext.}$ = −(0.50 ± 0.09) log $R$ + (25.18 ± 0.10) with a correlation coefficient $r$ = −0.36 and a chance probability of $p$ = 1.47 × 10$^{-7}$; for quasars, log $L_{ext.}$ = −(0.42 ± 0.06) log $R$ + (26.30 ± 0.07) with a correlation coefficient $r$ = −0.35 and a chance probability of $p$ = 2.43 × 10$^{-12}$; for Seyferts, log $L_{ext.}$ = −(0.38 ± 0.45) log $R$ + (23.62 ± 0.39) with a correlation coefficient $r$ = −0.12 and a chance probability of $p$ = 40.02%; for galaxies, log $L_{ext.}$ = −(0.67 ± 0.06) log $R$ + (24.51 ± 0.08) with a correlation coefficient $r$ = −0.41 and a chance probability of $p$ ~ 0; for FRIs and FRIIs, log $L_{ext.}$ = −(0.54 ± 0.15) log $R$ + (24.00 ± 0.21) with a correlation coefficient $r$ = −0.41 and a chance probability of $p$ = 4.51 × 10$^{-4}$. We can see that there are anti-correlations between core-dominance parameter and extended luminosity for BL Lacs, quasars, galaxies and FRIs and FRIIs, but there is no linear correlation for Seyfert galaxies.

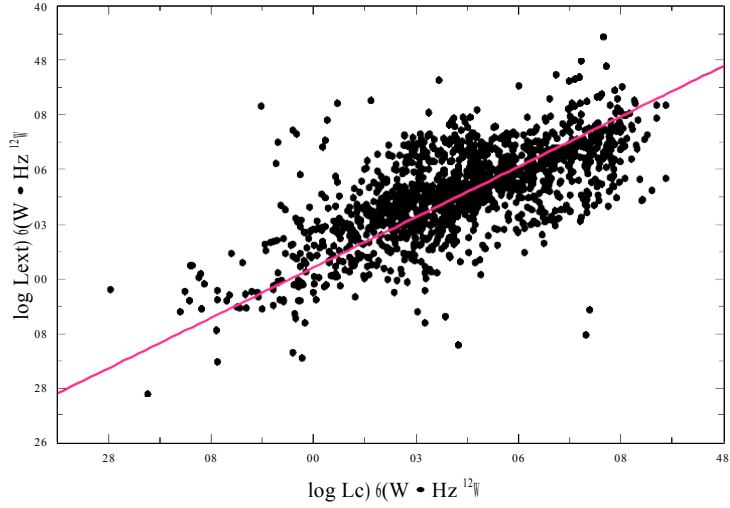

**Fig. 4** Plot of the extended luminosity (log $L_{ext.}$) against the core luminosity (log $L_{core}$) for whole sample. Linear fitting gives that log $L_{ext.}$ = (0.55±0.02) log $L_{core}$ +(11.70±0.53) ($r$ = 0.67, $p$ ~ 0).

**Table 6** Linear regression ($y = ax + b$) results for extended luminosity (log $L_{ext.}$) against core-dominance parameter (log $R$) in the whole sample

| Sample | ($a \pm \Delta a$) | ($b \pm \Delta b$) | $r$ | $p$ |
|---|---|---|---|---|
| Total | −0.30 ± 0.04 | 25.15 ± 0.05 | −0.21 | $p = 2.55 \times 10^{-15}$ |
| BL Lac | −0.50 ± 0.09 | 25.18 ± 0.10 | −0.36 | $p = 1.47 \times 10^{-7}$ |
| Quasar | −0.42 ± 0.06 | 26.30 ± 0.07 | −0.35 | $p = 2.43 \times 10^{-12}$ |
| Seyfert | −0.38 ± 0.45 | 23.62 ± 0.39 | −0.12 | $p = 40.02\%$ |
| Galaxy | −0.67 ± 0.06 | 24.51 ± 0.08 | −0.41 | $p$ ~ 0 |
| FRI & FRII | −0.54 ± 0.15 | 24.00 ± 0.21 | −0.41 | $p = 4.51 \times 10^{-4}$ |

It now investigates the fact that FRI radio galaxies are the parent population of BL Lacs while FRIIs are the parent population of quasars by using the correlation between extended luminosity and core-dominance parameter. On account of the radio galaxies have a large viewing angle, their beaming effect is very weak (Jorstad et al.2005;Xue et al.2017). Therefore the log $R$ of radio galaxies is very small. In this case, the beamed sources and their parent population should follow the same correlation. When we study the relationship between the log $L_{ext.}$ and log $R$ for BL Lacs ~ FR type I galaxies and quasars ~ FR type II galaxies respectively, we have log $L_{ext.} = -(0.22 \pm 0.54) \log R + (24.77 \pm 0.67)$ with a correlation coefficient $r = -0.15$ and a chance probability of $p = 3.79 \times 10^{-8}$ for the sample included BL Lacs and FR I galaxies; log $L_{ext.} = -(0.33 \pm 0.05) \log R + (26.21 \pm 0.06)$ with a correlation coefficient $r = -0.31$ and a chance probability of $p = 2.54 \times 10^{-4}$ for quasars and FR II galaxies (see Figure 6(a) and (b)). All results are presented in Table 6.

When we considered the correlation between the core-dominance parameter ($\log R$) and redshift ($\log z$) for the sources whose measured redshift is available (429 sources have no measured redshift) and we can obtain that $\log R = (0.22 \pm 0.05) \log z - (0.20 \pm 0.04)$ with a correlation coefficient $r = 0.21$ and a chance probability of $p = 1.89 \times 10^{-15}$ (see Figure 7).

The above correlation may arise from a selection effect. Sources with high redshift should be strongly beamed, dominating their extended emission resulting in a large core-dominance parameter, while for sources with low redshift have the emission from the core and from the extended parts can be detected. This can result in the fact that higher redshift sources have a larger core-dominance parameter.

## 3 DISCUSSION

Based on the two component model (Urry & Shafer 1984), the total observed flux density, $S^{ob}$, is the sum of the unbeamed $S_{unb}$ and the beamed components $S^{ob}_j$: $S^{ob} = S_{unb} + S^{ob}_j$. If we assume that the radio spectral index for the beamed, unbeamed and total emission are $\alpha_j$, $\alpha_{unb}$ and $\alpha_{total}$, receptively. We thus have $S^{ob}_j = S^{ob}_{j,0} \nu^{-\alpha_j}$, $S_{unb} = S_{unb,0} \nu^{-\alpha_{unb}}$ and $S^{ob}_o = S^{ob}_o \nu^{-\alpha_{total}}$ since the exponential spectrum, substituting,

$$S^{ob}_o \nu^{-\alpha_{total}} = S^{ob}_{j,0} \nu^{-\alpha_j} + S_{unb,0} \nu^{-\alpha_{unb}}, \quad (3)$$

then differentiating it with respect to the spectral index $\alpha$,

$$\alpha_{total} \cdot S^{ob}_o \nu^{-\alpha_{total}} = \alpha_j \cdot S^{ob}_{j,0} \nu^{-\alpha_j} + \alpha_{unb} \cdot S_{unb,0} \nu^{-\alpha_{unb}}, \quad (4)$$

adopting the definition of core-dominance parameter $R = \dfrac{S^{ob}_j}{S_{unb}}$ we obtained (Fan et al. 2010, 2011),

$$\alpha_{total} = \frac{R}{1+R}\alpha_j + \frac{1}{1+R}\alpha_{unb}, \quad (5)$$

we take consideration by $\alpha_j = \alpha_{core}$ and $\alpha_{unb} = \alpha_{ext.}$. Therefore,

$$\alpha_{total} = \frac{R}{1+R}\alpha_{core} + \frac{1}{1+R}\alpha_{ext.}, \quad (6)$$

the radio spectral index can be calculated as $\alpha_i = -\frac{\log (S^i_1/S^i_2)}{\log (\nu_2/\nu_1)}$, $i$ means total, core and extended, $S^i_1$ and $S^i_2$ are the emission in the frequency of $\nu_1$ and $\nu_2$ respectively. In our calculation, we consider that the spectral indices ($\alpha$) that we calculated above, equal to the total components of spectral indices for the sources, that is, $\alpha = \alpha_{total}$. We adopt this relationship to the current sample and study the correlation between core-dominance parameter log $R$ and the total radio spectral index $\alpha$, and obtain the theoretical fitting results of $\alpha_{core}$ and $\alpha_{ext.}$ (also see Fan et al. 2010, 2011).

When the radio spectral indices and core-dominance parameters from Table 1 are plotted in Figure 8, we can see a clear trend for the radio spectral index $\alpha$ to be a function of the core-dominance parameter. Figure

**Table 7** Theoretical fitting results for radio spectral index against core-dominance parameter in the whole sample

| Sample | $\alpha_{\rm core}$ | $\alpha_{\rm ext.}$ | $R^2$ | $p$ |
|---|---|---|---|---|
| Total | $-0.08 \pm 0.03$ | $1.04 \pm 0.05$ | 0.22 | $< 0.01$ |
| BL Lac | $-0.02 \pm 0.12$ | $0.70 \pm 0.22$ | 0.01 | $< 0.10$ |
| Quasar | $-0.34 \pm 0.08$ | $0.60 \pm 0.11$ | 0.04 | $< 0.05$ |
| Galaxy | $-0.05 \pm 0.07$ | $0.88 \pm 0.03$ | 0.21 | $< 0.01$ |
| FRI & FRII | $-0.12 \pm 0.22$ | $1.04 \pm 0.07$ | 0.13 | $< 0.05$ |
| Seyfert | ... | ... | ... | ... |

8 implies that we cannot use one simple curve to fit all points because the radio spectral indexes, $\alpha_{\rm core}$ and $\alpha_{\rm ext.}$, are different for different sources. One possibility is that the flux densities used to calculate the spectral index and those used to calculate core-dominance are not simultaneous, which results in scattering points (see Fan et al.2011). If all of the sources follow relation (6), then we can estimate the spectral indices, $\alpha_{\rm core}$ and $\alpha_{\rm ext.}$, for the whole sample by minimizing $\Sigma[\alpha_{\rm total} - \alpha_{\rm core} R/(1 + R) + \alpha_{\rm ext.}(1 + R)]^2$. When we adopted this across the whole sample, $\alpha_{\rm core} = -0.08 \pm 0.03$ and $\alpha_{\rm ext.} = 1.04 \pm 0.05$ were obtained ($R^2 = 0.22$, $p < 0.01$). The fitting result is shown in the curve fit in Figure 8. The fitting results are consistent with the general consideration taking $\alpha_{\rm core} = 0$ and $\alpha_{\rm ext.} = 0.75$ (Fan et al.2011, also see Pei et al.2016).

When we considered the subclasses separately, we can obtain a plot of the spectral index against the core-dominance parameter as shown in Figure 9(a) to (d) for BL Lacs, quasars, galaxies and FRIs & FRIIs, respectively. The fitting results give that $\alpha_{\rm core} = -0.02 \pm 0.12$ and $\alpha_{\rm ext.} = 0.70 \pm 0.22$ ($R^2 = 0.01$, $p < 0.10$); $\alpha_{\rm core} = -0.34 \pm 0.08$ and $\alpha_{\rm ext.} = 0.60 \pm 0.11$ ($R^2 = 0.04$, $p < 0.05$); $\alpha_{\rm core} = -0.05 \pm 0.07$ and $\alpha_{\rm ext.} = 0.88 \pm 0.03$ ($R^2 = 0.21$, $p < 0.01$); $\alpha_{\rm core} = -0.12 \pm 0.22$ and $\alpha_{\rm ext.} = 1.04 \pm 0.07$ ($R^2 = 0.13$, $p < 0.05$), respectively. However, for Seyfert galaxies, we cannot get an appropriate fitting. All results are given in Table 7. The tendency for the spectral index to depend on the core-dominance parameter is probably due to the relativistic beaming effect and our fitting results imply that different subclasses show different degrees of relevance with regard to the beaming effect.

We apply the relation (6) to the our sample and study the correlation between core-dominance parameter $\log R$ and total radio spectral index $\alpha$, and obtain the theoretical fitting results of $\alpha_{\rm core}$ and $\alpha_{\rm ext.}$. However, observations are not all available for the considered frequency (it is 5 GHz in the present work), therefore, we have to transfer the observed values at other frequency into the considered frequency (5 GHz). From the previous studies, some authors have calculated the radio spectral indices in the core and extended components and obtain the average values as $\alpha_{\rm core} \sim 0$ and $\alpha_{\rm ext.} \sim 0.75$. So, we apply these two values to our calculation of core-dominance parameters if $\alpha_{\rm core}$ and $\alpha_{\rm ext.}$ are not known. Using relation (6), we can get fitting results of $\alpha_{\rm core}$ and $\alpha_{\rm ext.}$ for a group.

In this paper, our results confirmed the conclusion of Fan et al.(2011), who obtained that $\log R|_{\rm BL\ Lac} > \log R|_{\rm quasar} > \log R|_{\rm Seyfert} > \log R|_{\rm galaxy} > \log R|_{\rm FRI} > \log R|_{\rm FRII}$ averagely, and also roughly consistent with the distributions of their beaming factor values (e.g., Jorstad et al.2005; Richards & Lister 2015; Sun et al.2015; Xue et al.2017). Does this is the conclusion of universality for AGNs?

From the previous studies, the core-dominance parameter $R$ can take the role of the indicator of beaming effect (see Urry & Padovani 1995, Fan 2003),

$$R(\theta) = f\gamma^{-n}[(1-\beta\cos\varphi)^{-n+\alpha} + (1+\beta\cos\varphi)^{-n+\alpha}], \qquad (7)$$

where $f$ is the intrinsic ratio, defined by the intrinsic flux density in the jet to the extend flux density in the co-moving frame, $f = \frac{S_{\text{core}}^{\text{in}}}{S_{\text{ext.}}^{\text{in}}}$ (Fan & Zhang 2003), $\theta$ the viewing angle, $\gamma$ the Lorentz factor, $\gamma = (1-\beta^2)^{-1/2}$, $\alpha$ is the radio spectral index and $n$ depends on the shape of the emitted spectrum and the physical detail of the jet, $n = 2$ for continuous jet and $n = 3$ for blobs. Therefore, $R$ is a good statistical measure and indicator of the relativistic beaming effect.

Extragalactic sources show very strong radio emission. If we consider the two component model of radio emission, which is composed of a compact relativistically beamed core component and an unbeamed lobe component. The core-dominance parameter defined as $R = S_{\text{core}}/S_{\text{ext.}}$ is taken as an orientation parameter and therefore the indicator as beaming effect. Ghisellini et al. (1993) compiled a sample of extragalactic sources and obtained that the average of ($\log R$) of BL Lacs is larger than that in quasars. Fan et al. (2011) also found this tendency. However, Murphy et al. (1993) obtained a distribution of ($\log R$) for a sample of 74 blazars in 5 GHz and found there is a tendency for BL Lacs to have lower ($\log R$) than quasars. We think the reason is that Murphy et al. (1993) only considered the bright sources ($S_{5\text{ GHz}} > 1$ Jy) while the extragalactic sources considered by Ghisellini et al. (1993), Fan et al. (2011) and this paper runs a broad range of flux densities.

Previously many researchers have studied the unification for BL Lacs & FRIs and quasars & FRIIs (see Urry et al. 1991; Xie et al. 1993; Fan et al. 2011; Xue et al. 2017), they all argue that BL Lacs are unified with FRIs while quasars with FRIIs. Similar results are also obtained based on infrared (Fan et al. 1997) and X-ray studies (Wang et al. 2006b). Fanaroff & Riley (1974) defined FR type I galaxy and type II galaxy as using the ratio of the distance between the regions of highest brightness on the opposite side of the central galaxy to the total extent of the source measured, classified in type I when this ratio is less than 0.5, while in type II when greater than 0.5. Therefore, as the definition of core-dominance parameter $R$, we can expect that $(\log R)|_{\text{FRI}}$ is higher than $(\log R)|_{\text{FRII}}$. In the popular unification scenario by relativistic beaming effect, BL Lacs are believed to be the beamed counterparts of FRI radio galaxies, while quasars are believed to be the beamed counterparts of FRII radio galaxies (Urry & Padovani 1995). The $\log R$ for BL Lacs and quasars is quite large ($R \gg 1$), which shows that the core emission in BL Lacs and quasars are expected to be relativistic beamed. On the other hand, the $\log R$ for FRIs and FRIIs are very small ($R \ll 1$). Perhaps it shows that the beaming effect in FRIs and FRIIs is less important. From our discussion, BL Lacs have the same range of extended luminosity as the FRIs, while quasars have the same range as the FRII (see Figure 3). Many authors studied the "unified scheme" as BL Lacs ~ FRIs and quasars ~ FRIIs (Ubachukwu & Chukwude 2002; Fan et al. 2011; Odo & Ubachukwu 2013; Odo et al. 2017), they believed FRIs are expected to be the parent population of the BL Lacs, while FRIIs expected to be the parent population of the quasars. In this paper, the K-S test results of the extended luminosity indicate that the null hypothesis (they both are from the same population) cannot be rejected at the following confidence level for the different samples: $p = 27.5\%$ with $d_{\max} = 0.16$ for BL Lacs-FRIs, $p = 9.03\%$ with $d_{\max} = 0.17$ for quasars-FRIIs.

Therefore, we confirm the agreement with the BL Lacs ~ FRIs and quasars ~ FRIIs unified scheme by the study of the extended luminosity (log $L_{ext.}$), which shows that the unification scheme for them cannot be rejected. In this sense, we can claim that BL Lacs ~ FRIs and quasar ~ FRIIs are homogeneous class of the AGNs.

In the plot of extended luminosity against core-dominance parameter (Figure 5), we can see that there is a trend for the extended luminosity to be anti-correlated with the core-dominance parameter. In a two-component model, $R = L_{core}/L_{ext.}$, which can be expressed in the form

$$1 + R = \frac{L_{core} + L_{ext.}}{L_{ext.}} = \frac{L_{total}}{L_{ext.}}, \tag{8}$$

Figure 10 shows that the distribution of total luminosity, $\log L_{total}$ (W · Hz$^{-1}$), for the whole sources. The Gaussian fitting gives that $\mu = 25.94 \pm 0.04$ and $\sigma = 0.65 \pm 0.04$ with $R^2 = 0.96$ and $p = 4.27 \times 10^{-12}$. From this result, we can assume that if the total luminosity of our sample, $L_{total} = L_{core} + L_{ext.}$, is a constant, then $R + 1$ is anti-correlated to $L_{ext.}$. This correlation implies that a lower extended luminosity source has a larger core-dominance parameter, $R$, implying either a large $f$ or a large $\delta$ since $R \propto f\delta^p$ (Fan et al. 2011).

We obtain negative correlations for each different kind of AGNs between $\log L_{ext.}$ and $\log R$. If both $L_{ext.}$ and $L_{core}$ have a distribution with upper and lower limits, then it is easier to get a large $R$ for a source with $L_{ext.}$ close to the lower luminosity limit, and vice versa. Thus, sources with larger $\delta$ are brighter, and we get an anti-correlation between the core-dominance parameter and the extended luminosity (see Figure 5).

The association between the spectral index $\alpha$ and core-dominance parameter $R$ in radio sources is a subject of further study. Fan et al. (2011) calculated the core-dominance parameter and the radio spectral index for the whole sample, and gave the relationship between $\alpha$ and $\log R$, which indicates the spectral index is associated with core-dominance parameter. We also suggest that the relativistic beaming effect may result in an association between spectral index and core-dominance parameter for extragalactic sources in radio emission (also see Pei et al. 2016). In the two component beaming model, the relative prominence of the core with respect to the extended emission defined as the ratio of core-to-extended-flux density measured in the rest frame of the source $\log R$ has become a suitable statistical measure of beaming and orientation.

In our previous work (Fan et al. 2011), we collected 1223 AGNs, including 77 BL Lacs, 495 quasars, 180 Seyfert galaxies, 280 galaxies, 119 FRIs & FRIIs and 72 unidentified sources, and calculating their core-dominance parameters ($\log R$) and radio spectral indices ($\alpha$). Our results show that the distributions of $\log R$ and $\alpha$ in different subclasses are different with $\log R|_{BL\ Lac} > \log R|_{quasar} > \log R|_{galaxy/Seyfert} > \log R|_{FRI/II}$, while $\alpha|_{BL\ Lac} < \alpha|_{quasar} < \alpha|_{galaxy/Seyfert} < \alpha|_{FRI/II}$ on average. In this paper, we enlarge the AGNs sample, which contains 250 BL Lacs, 520 quasars, 175 Seyferts, 1178 galaxies, 153 FRIs & FRIIs and 104 unidentified sources. The comparison of previous work and this paper is shown in Table 8. From this Table, we can see that the average value for core-dominance parameter ($\langle \log R \rangle$) for the whole AGNs sample in Fan et al. (2011) and this work is similar: $\langle \log R \rangle|_{Total} = -0.35$ for Fan et al. (2011) and $\langle \log R \rangle|_{Total} = -0.34$ for this work. As for sub-classes, we have $\langle \log R \rangle|_{BL\ Lac} = 0.87, 0.55$; $\langle \log R \rangle|_{quasar} = 0.13, 0.24$; $\langle \log R \rangle|_{Seyfert} = -0.39, -0.37$; $\langle \log R \rangle|_{galaxy} = -0.93, -0.67$; $\langle \log R \rangle|_{FRI} = -1.41, -0.97$ and $\langle \log R \rangle|_{FRII} = -2.09, -1.38$ for

Table 8 Comparison of statistical results between Fan et al.(2011) and this work

| Sample | N | | (log R) | | (α) | | α$_{core}$ | | α$_{ext.}$ | |
|---|---|---|---|---|---|---|---|---|---|---|
| | Fan11 | TW | Fan11 | TW | Fan11 | TW | Fan11 | TW | Fan11 | TW |
| Total | 1223 | 2400 | -0.35 | -0.34 | 0.51 | 0.41 | -0.07 | -0.08 | 0.92 | 1.04 |
| BL Lac | 77 | 250 | 0.87 | 0.55 | 0.16 | 0.22 | -0.01 | -0.02 | 0.65 | 0.70 |
| Quasar | 495 | 520 | 0.13 | 0.24 | 0.36 | 0.15 | -0.09 | -0.34 | 0.89 | 0.60 |
| Seyfert | 180 | 175 | -0.39 | -0.37 | 0.53 | 0.43 | -0.01 | ⋯ | 0.91 | ⋯ |
| Galaxy | 280 | 1178 | -0.93 | -0.67 | 0.73 | 0.57 | -0.01 | -0.05 | 0.91 | 0.88 |
| FRI | 18 | 46 | -1.41 | -0.97 | 0.80 | 0.69 | 0.34 | -0.12 | 0.97 | 1.04 |
| FRII | 101 | 107 | -2.09 | -1.38 | 0.96 | 0.60 | 0.34 | -0.12 | 0.97 | 1.04 |

Notes: Here, 'Fan11' refers to Fan et al.(2011) and 'TW' to this work.

Fan et al.(2011) and this paper, respectively. Besides, we also discuss the relation between core-dominance parameter (log $R$) and radio spectral index ($\alpha$) (see relation (6)), as the results in Fan et al.(2011) (see Table 8), we obtained the similar relation and got the fitting values for $\alpha_{core}$ and $\alpha_{ext.}$ as well.

As the observation tools and observation methods continue to advance and the observational accuracy becomes higher and higher, the division of the core regions will become clearer and clearer and the scale of the extended region is getting larger and larger, and therefore, the extended flux value will also increase. Since the core emission is strongly beamed with the jet pointing to the sight line of the observer and dominates the emission, the flux decreases caused by the decrease of the core size. In this sense, according to the definition of the core-dominance parameter ($R = S_{core}/S_{ext.}$), we can expect that $R$ will become smaller. Therefore, we think that this kind of consideration will not affect our discussion of the principal facts.

For blazars, (log $R$) is greater than 0, which indicates that the core component for blazars dominates the emission. Secondly, the tendency for core-dominance parameters (log $R$) and spectral indices ($\alpha$) are opposite with blazars taking the maximum value for (log $R$) while the minimum for ($\alpha$). Third, we further confirm that the two-component model in emission and there is a relation between log $R$ and $\alpha$. The fitting results for $\alpha_{core}$ and $\alpha_{ext.}$ are also given.

According to relation (6), in the relativistic beaming scenario for highly beamed sources, we have log $R$ ≫ 0, which leads to $\alpha_{total} \approx \alpha_{core}$. And $\alpha$ is dominated by $\alpha_{ext.}$ for the case log $R$ ≪ 0. Thus, to use this relation as a test for relativistic beaming, we require sources with log $R$ > 0. Therefore, we can consider that the association between core-dominance parameters and spectral indices may suggest that relativistic beaming could influence the spectral characteristics of this extreme class of objects.

## 4 CONCLUSION

From our discussions, given the the core-dominance parameter, log $R$ and radio spectral index, $\alpha$, the $\alpha_{core}$ and $\alpha_{ext.}$ can be obtained. In this paper, we compiled 2400 objects with the relevant data to calculate the core-dominance parameters. The sample is still not sufficient enough, but we think it is adequate for statistical analysis. Therefore, we can draw the following conclusion:

1. Core-dominance parameters ($\log R$) are quite different for different subclasses of AGNs: on average, the following sequence holds: $\log R|_{\text{BL Lac}} > \log R|_{\text{quasar}} > \log R|_{\text{Seyfert}} > \log R|_{\text{galaxy}} > \log R|_{\text{FRI}} > \log R|_{\text{FRII}}$.

2. A theoretical correlation fitting between core-dominance parameter ($\log R$) and radio spectral index ($\alpha$) is adopted and also obtained for all subclasses, which means the spectral index is dependent on the core-dominance parameter, probably from the relativistic beaming effect. And $\alpha_{\text{core}} = -0.08$ and $\alpha_{\text{ext.}} = 1.04$ are obtained.

3. There is an anti-correlation between extended-luminosity ($\log L_{\text{ext.}}$) and core-dominance parameter in different kinds of objects.

4. Sources with larger redshift show a larger core-dominance parameter.

5. BL Lac objects and FRI radio galaxies are unified, while quasars and FRII radio galaxies are unified.

**Acknowledgements** We thank the anonymous referee for the constructive comments and suggestions. This work is partially supported by the National Natural Science Foundation of China (11733001, U1531245, NSFC 10633010, NSFC 11173009), Natural Science Foundation of Guangdong Province (2017A030313011), and supports for Astrophysics Key Subjects of Guangdong Province and Guangzhou City.

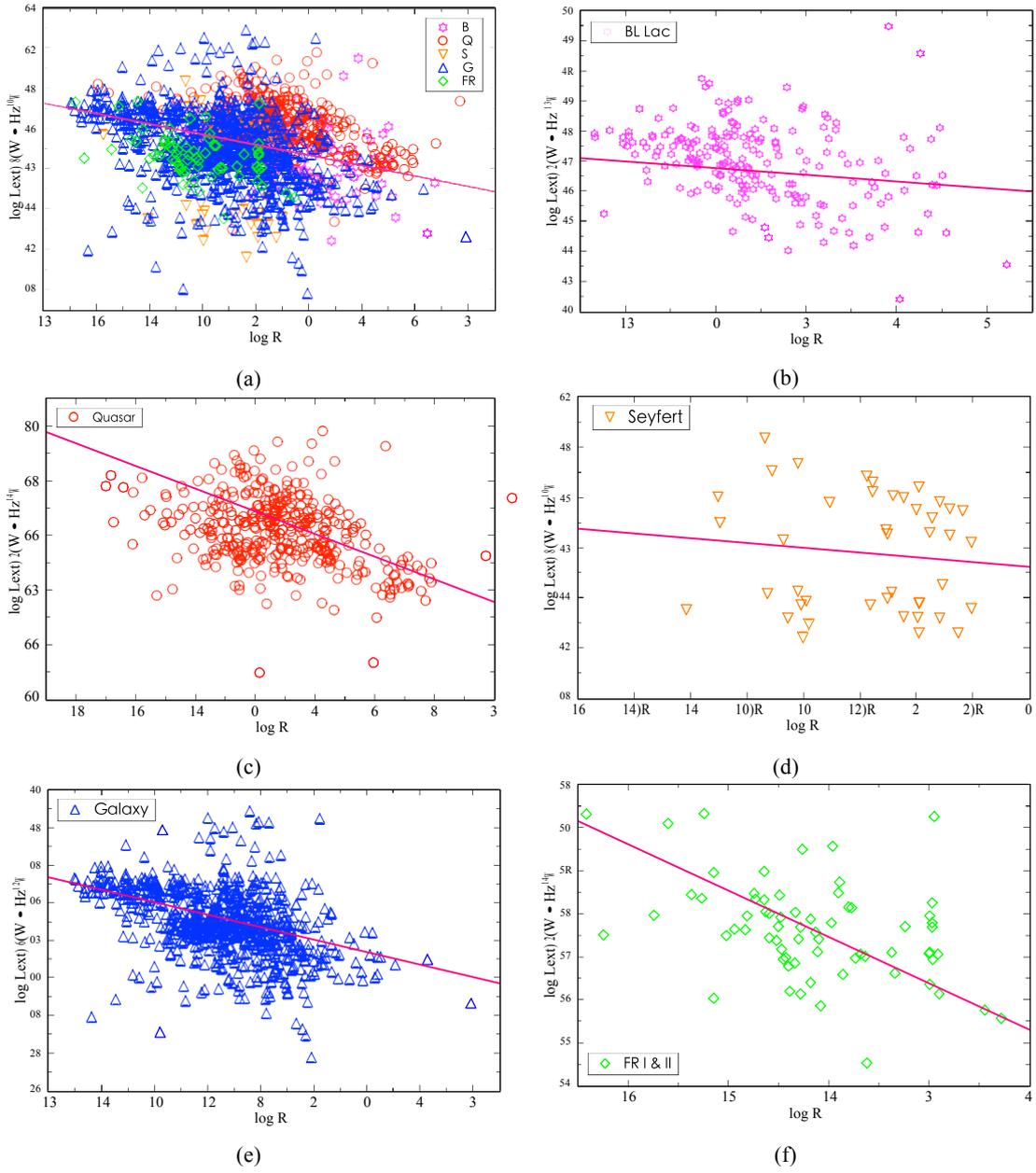

**Fig. 5** Plot of the extended luminosity (log $L_{ext.}$) against the core-dominance parameter (log $R$) for (a) whole sample; (b) BL Lacs; (c) quasars; (d) Seyferts; (e) galaxies and (f) FR type Is and IIs. The linear regression gives that (a) log $L_{ext.}$ = −(0.30±0.04) log $R$ + (25.15±0.05) ($r$ = −0.21, $p$ = 2.55 × 10$^{-15}$); (b) log $L_{ext.}$ = −(0.50 ± 0.09) log $R$ + (25.18 ± 0.10) ($r$ = −0.36, $p$ = 1.47×10$^{-7}$); (c) log $L_{ext.}$ = −(0.42±0.06) log $R$+(26.30±0.07) ($r$ = −0.35, $p$ = 2.43 × 10$^{-12}$); (d) log $L_{ext.}$ = −(0.38 ± 0.45) log $R$ + (23.62 ± 0.39) ($r$ = −0.12, $p$ = 40.02%); (e) log $L_{ext.}$ = −(0.67±0.06)$logR$ + (24.51±0.08) ($r$ = −0.41, $p$ ~ 0) and (f) log $L_{ext.}$ = −(0.54±0.15) log $R$ + (24.00±0.21) ($r$ = −0.41, $p$ = 4.51×10$^{-4}$). In this plot, the representations of all symbols are the same as in Figure 1.

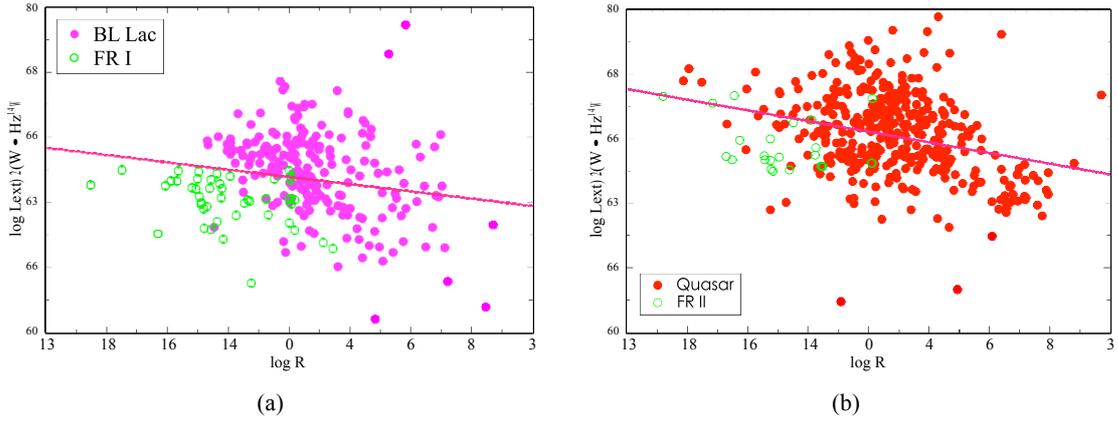

(a)        (b)

**Fig. 6** Plot of the extended luminosity ($\log L_{\text{ext.}}$) against the core-dominance parameter ($\log R$) for the sample of (a) BL Lacs and FR type I galaxies and (b) quasars and FR type II galaxies. Linear fitting gives that (a) $\log L_{\text{ext.}} = -(0.22 \pm 0.54)\log R + (24.77 \pm 0.67)$ ($r = -0.15$, $p = 3.79 \times 10^{-8}$) and (b) $\log L_{\text{ext.}} = -(0.33 \pm 0.05)\log R + (26.21 \pm 0.06)$ ($r = -0.31$, $p = 2.54 \times 10^{-4}$). Here, (a) magenta solid circles stand for BL Lacs while green hollow circles stand for FR type I galaxies and (b) red solid circles stand for quasars while green hollow circles stand for FR type II galaxies.

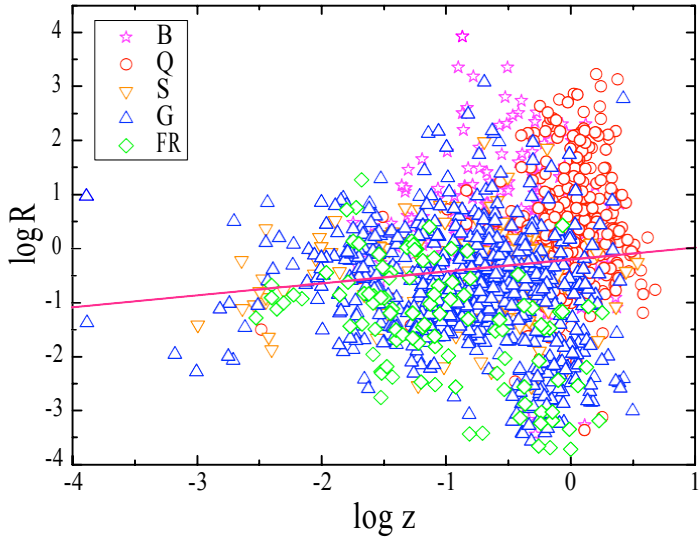

**Fig. 7** Plot of the redshift ($\log z$) against the core-dominance parameter ($\log R$) for the whole sample. Linear fitting gives that $\log R = (0.22 \pm 0.05)\log z - (0.20 \pm 0.04)$ ($r = 0.21$, $p = 1.89 \times 10^{-15}$). Here, the representations of all symbols are the same as in Figure 1.

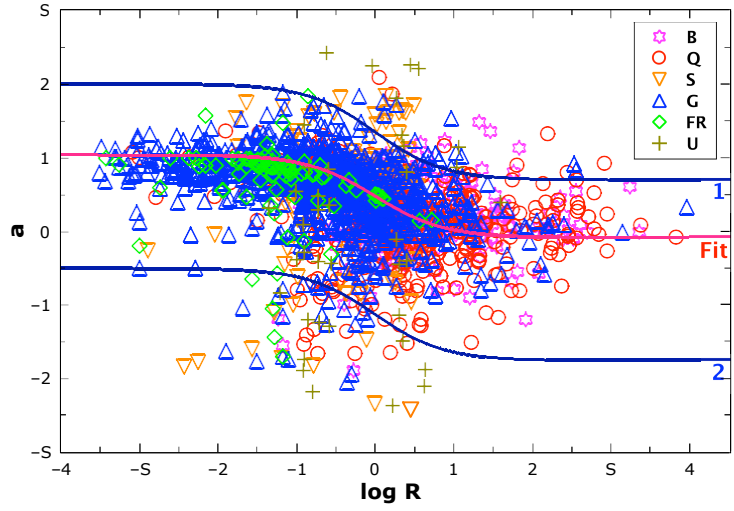

**Fig. 8** Plot of the the radio spectral index, $\alpha_R$, against the core-dominance parameter, $\log R$, for the whole sample. In this plot, magenta ✦ stands for BL Lac, red ◯ for quasar, orange ▽ for Seyfert, blue △ for galaxy, green ◇ for FRI & FRII, yellow dark + for unidentified sources. Curve 1 corresponds to $\alpha_{core}$ (or $\alpha_j$)= 0.70 and $\alpha_{ext.}$ (or $\alpha_{unb}$)= 2.00, curve 2 corresponds to $\alpha_{core}$ = −1.75 and $\alpha_{ext.}$ = 0.50, and the fitting curve corresponds to $\alpha_{core}$ = −0.08 and $\alpha_{ext.}$ =1.04.

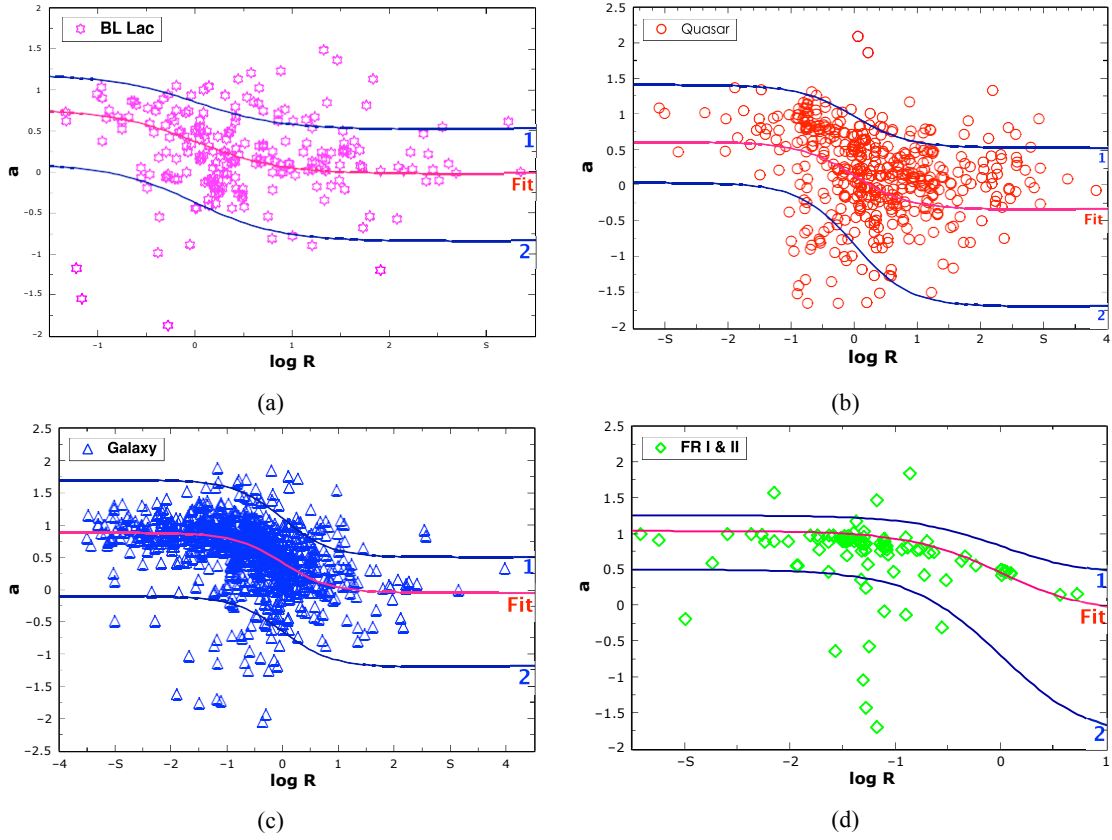

**Fig. 9** Plot of the radio spectral index, $\alpha_R$, against the core-dominance parameter, $\log R$, for (a) BL Lacs. Curve 1 corresponds to $\alpha_{core}$ (or $\alpha_j$)= 0.52 and $\alpha_{ext.}$ (or $\alpha_{unb}$)= 1.18, curve 2 corresponds to $\alpha_{core}$ = −0.84 and $\alpha_{ext.}$ = 0.10, and the fitting curve corresponds to $\alpha_{core}$ = −0.02 and $\alpha_{ext.}$ = 0.70; (b) quasars. Curve 1 corresponds to $\alpha_{core}$ (or $\alpha_j$)= 0.52 and $\alpha_{ext.}$ (or $\alpha_{unb}$)= 1.40, curve 2 corresponds to $\alpha_{core}$ = −1.70 and $\alpha_{ext.}$ = 0.03, and the fitting curve corresponds to $\alpha_{core}$ = −0.34 and $\alpha_{ext.}$ = 0.60; (c) galaxies. Curve 1 corresponds to $\alpha_{core}$ (or $\alpha_j$)= 0.50 and $\alpha_{ext.}$ (or $\alpha_{unb}$)= 1.70, curve 2 corresponds to $\alpha_{core}$ = −1.20 and $\alpha_{ext.}$ = −0.10, and the fitting curve corresponds to $\alpha_{core}$ = −0.05 and $\alpha_{ext.}$ = 0.88; (d) FRIs & FRIIs. Curve 1 corresponds to $\alpha_{core}$ (or $\alpha_j$)= 0.40 and $\alpha_{ext.}$ (or $\alpha_{unb}$)= 1.26, curve 2 corresponds to $\alpha_{core}$ = −1.90 and $\alpha_{ext.}$ = 0.50, and the fitting curve corresponds to $\alpha_{core}$ = −0.12 and $\alpha_{ext.}$ = 1.04.

24            Pei

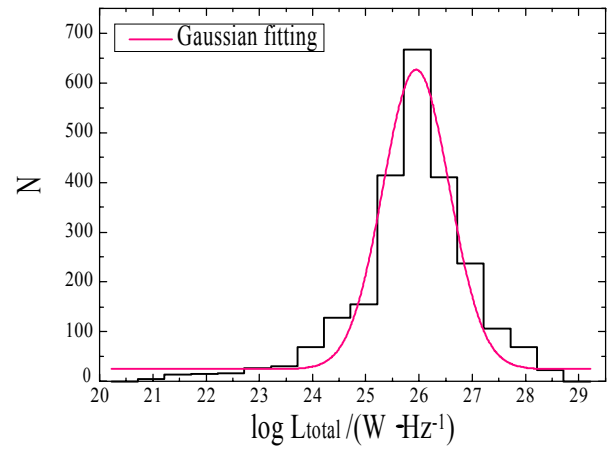

**Fig. 10** Distribution of total luminosity, $\log L_{\text{total}}$, for the whole sources. The Gaussian fitting gives that $\mu = 25.94 \pm 0.04$ and $\sigma = 0.65 \pm 0.04$ with $R^2 = 0.96$ and $p = 4.27 \times 10^{-12}$.